\pdfoutput=1

\documentclass[aps,prl,10pt,twocolumn,showpacs,superscriptaddress,footnoteinbib]{revtex4}
\usepackage{amsmath}
\usepackage{latexsym}
\usepackage{amssymb}
\usepackage{graphics,epstopdf}

\begin{document}
 \title{ CTC assisted PR box type correlation can lead to signaling}
\author {Indranil Chakrabarty}
\affiliation{Institute of Physics, Sainik School Post, 
Bhubaneswar-751005, Orissa, India.\\
Centre of Security Theory and Algorithimic Research, IIIT Hyderabad,Gachibowli, Hyderabad-500032,
Andhra Pradesh, India.}
\author {Tanumoy Pramanik}
\affiliation{S. N. Bose National Centre for Basic Sciences, Salt Lake,
Kolkata-700 098, India }
\author {Arun K Pati}
\affiliation{Harish Chandra Research Institute, Chhatnag Road, Jhunsi, Allahabad-211019, UP, India}
\author{Pankaj Agrawal}
\affiliation{Institute of Physics, Sainik School Post, 
Bhubaneswar-751005, Orissa, India}
\date{\today}

\begin{abstract}
It is known that there exist non-local correlations that respect no-signaling
criterion, but violate Bell-type inequalities more than quantum-mechanical correlations.
Such super quantum correlations were introduced as the Popescu-Rohrlich (PR) box.
We consider such non-local boxes with
two/three inputs and two/three outputs. We show that these super quantum 
correlations can lead to signaling when at least one of the input bit has access to a 
word line along a closed time-like curve. 
\end{abstract}

\maketitle

\section{I. Introduction}
 One of the most intriguing features of quantum physics is the nature of
correlations present in the composite systems. These correlations can be captured by 
the amount of entanglement present in the system \cite{woo}. 
It is well known that the quantum mechanical correlations can violate Bell-type
inequalities, but are not strong enough to violate 
the causality. If we take the no-signaling principle as an upper bound to  all possible admissible
correlations that nature allows, then it is interesting to find out if there exist any such correlations that go 
beyond quantum limit without violating causality \cite{pop,cri}. Popescu and Rohrlich constructed a 
hypothetical black box (PR box) which exhibits 
correlations beyond the standard quantum limit in spite of being perfectly consistent with the
no-signaling criterion \cite{pop}. Recently, there have been growing interest in investigating 
the power of the PR boxes. For example, it has been shown \cite{bar} that any two-output bipartite
box can be simulated with the help of PR boxes.
 Furthermore, it turns out that there exists multiparty correlations
which cannot be simulated by using $n$ PR boxes for arbitrary $n$ \cite{bar}. Some authors have also tried 
to find quantum mechanical realization of the PR box by focusing on the relation between
causality and non-locality in the context of pre- and post selected
ensembles \cite{marco}.
% Researchers also strengthen the existing bound on the correlations in the 
%EPR-Bohm setup under certain assumption on the probabilities\cite{mat}. 
The authors in reference \cite{fit} have attempted to
quantify the amount of non-locality contained in the $n$ noisy versions of the PR boxes. Various other classifications according to  various definitions of multipartite non-locality and their
relation  to Bell-type inequalities have been studied \cite{piro,bar1}.
Two-party Gaussian states \cite{tas} have been used to probe experimentally
non-local Popescu-Rohrlich correlations. A maximum violation of the
CHSH inequality of $3.42$ is obtained. This actually  corresponds to the implementation 
of a non-local AND gate with success probability of $0.93$.
The PR boxes  provide great power and resources to carry out information-theoretical tasks, but 
they cannot produce all types of multiparty correlations.
It has also been seen that if one could clone a PR box, it could lead to signaling \cite{scar}. 
Quite analogous to  PR boxes, others have constructed hypothetical boxes like M boxes to
simulate the correlation of
 non-maximally entangled states in two-qubit systems \cite{ahanj,ahanj1}.
These non-local boxes were used to provide a winning strategy for the impossible 
coloring pseudo-telepathy game for the set of vectors having Kochen-Specker property in four 
dimension \cite{kunkri1}.\\

%Among its features, the ???anomaly of non-locality??? and a great power for
%information-theoretical tasks. Among its failures, the impossibility of reproducing all multi-partite
%distributions and the triviality of the allowed dynamics.\\
%There can be multiple interpretion associated with the the idea of non locality.
%However for our convenience we choose the following definition of nonlocality.\\

 In recent years, the power of a qubit moving on a closed time-like curve (CTC)
 have been explored in quantum information science. The existence of
closed time-like curve  is sometimes considered as an ingredient of a science fiction 
story in spite of being a theoretical possibility. A closed time-like curve is like a loop which typically
connects back to itself, through a  space time wormhole thus by linking a future space time
point with a past space time point. However, the existence of such world lines
lead to  paradoxes like `grandfather paradox.' Some years ago, David Deutsch proposed
a model within the quantum computational premise \cite{deu} which provides a self consistent 
solution for the interaction of a system moving on a CTC world line with a system 
following a casual world line. It has
also been speculated that in the presence of CTC, certain weird phenomenons like
enhancement of the computational power \cite{bru,bac,aar}, perfect cloning \cite{ahn}, distinguishabilty
of non-orthogonal states can take place \cite{bru1}. Also, it has been proved that if 
a density operator describes a CTC system and interacts with a CR system in a
 consistent way then that has to be a `proper' mixture \cite{pati}. However,  some authors have a opposite 
view of this and  have proposed a  different model of
 CTC (post selected CTC) which they claim to be free
from the drawbacks of the Deutsch model \cite{ben1}. \\ 

In the presence of a CTC, the Deutsch formalism describes
interaction between   a causality-respecting (CR) quantum system 
with another system that has a CTC world line. The states of these 
systems are described
by density matrices, though  there are restrictions on purifying a 
density matrix in CTC theory \cite{pati}. In this formalism, the CR system and
the CTC system evolve unitarily, ie., $\rho_{\rm CR}\otimes \rho_{\rm CTC} \rightarrow
(U \rho_{\rm CR}\otimes \rho_{\rm CTC} U^{\dagger})$.
For each initial mixed state $\rho_{CR}$ of the CR system, there exists a CTC system 
$\rho_{\rm CTC}$ such that we must have (the self-consistency condition) 
\begin{equation}
{\rm  Tr}_{\rm CR}(U \rho_{\rm CR}\otimes \rho_{\rm CTC} U^{\dagger})=\rho_{\rm CTC}.
\end{equation}
Mathematically, the solution to this equation is a fixed point. The final state of the CR system is 
then defined as
\begin{equation}
{\rm Tr}_{\rm CTC}(U \rho_{\rm CR}\otimes \rho_{\rm CTC} U^{\dagger}) = 
\rho_{\rm CR}^{'}.
\end{equation}
These are the basic two conditions which govern
the unitary dynamics of the CTC and CR quantum systems in the Deutsch model. \\

In this work, we consider the genuine two party and three-party non-local  boxes
 where  part of the box has access to a world line along a closed time-like curve. Interestingly,
  we find that if inputs of a 
PR box is a system with a CTC word line, then there is a violation of the no-signaling principle. 
In other words, in the presence of a CTC, these no-signaling
PR boxes can be converted into the boxes exhibiting signaling. Thus, the CTC assisted PR boxes can be 
called as the signaling boxes. Similar phenomenon can be observed for Mermin and Svetlichny boxes. \\

The organization of the paper is as follows. In section II, we consider two-party PR box 
and show that it leads to signaling when one of the 
inputs has CTC world line. In section III we study three-party non-local boxes like Sevlitchny and 
Mermin boxes with some of their inputs as CTC system. Finally, we conclude in the last section.

\section{II. Bipartite case}

The non-local boxes showing super quantum correlations have an interesting feature that
they violate the Criselon bound \cite{cri} in spite of being totally consistent with 
the no-signaling criterion. The box
 can be considered as a channel with two distinct inputs ($x$ for Alice and $y$ for Bob) 
and two distinct outputs 
($a$ for Alice and $b$ for Bob). Each of these inputs and outputs are bits and can assume the values 
$0$ and $1$ (all sums of two or more bits are taken as modulo 2).
The channel must satisfy the no-signaling condition. In other words 
the inputs and outputs on one side must be independent of inputs and outputs on the other side.
This is equivalent of saying that the marginals of Alice (Bob) do not depend on the input used 
by Bob (Alice). Mathematically, this means $\sum_{b=0,1} P(a,b|x,y) = P(a|x,y) =  P(a|x)$,
$\sum_{a=0,1} P(a,b|x,y) = P(b|x,y) = P(b|y)$.
Also, Alice's and Bob's marginals are the completely random distributions for both the 
values of the input, i.e., $P(a|x)=P(b|y)=\frac{1}{2}$.\\

 The PR box is defined in such a  way that Alice's and Bob's outcomes are perfectly correlated. 
The PR-box correlations (i.e., correlation between input and output) are given below
\begin{eqnarray}
a\oplus b=x.y,
\end{eqnarray}
where, `$x$' and `$y$' are the input of the PR-box and $a$ and $b$ are output of the PR-box.\\

In other words, the PR-correlation is no-signaling iff probability distribution of the output for given
input is 
\begin{eqnarray}
p(a,b|x,y) &=& \frac{1}{2} ~~~~~~\text{iff $a\oplus b=x.y$}, \nonumber \\
&=&0~~~~~~\text{otherwise}.
\end{eqnarray} 

 Explicitly, this means that $a=b$ holds when either $x=0$ or $y=0$ or both, while $a\neq b$ holds for $x=y=1$.
It can be seen that the no-signaling condition is satisfied. 
It tells us that the output at Alice's place for a given input by her should not depend on the input at the remote location (say Bob's place) and vice-versa. \\

Next, we consider  a situation where we begin with the same kind of a hypothetical PR box where one
of its input say on Bob's side is not a regular one; but it is a bit which has a
world line along a closed time-like curve. Since a classical bit $0$ or $1$ is equivalent 
to a qubit in one of the orthogonal states $|0\rangle$ or $|1\rangle$, a CTC bit can be thought 
of as a qubit in the state $|0\rangle$ or $|1\rangle$ traveling along a CTC. So whatever goes out of 
the box comes back in a loop making the
input and the output same. This is what the kinematic condition of the closed time-like curve demands.
When we apply Deutsch condition in this classical situation, mathematically, this is equivalent to the fact that the input
and the output on Bob's side are identical. For example, inn the reference \cite{bru}, it has been shown that in a completely classical situation like in the 
program of factoring large numbers, the Deutsch condition has been satisfied by considering a time register whose inputs and outputs are identical.
Here also we start with a box with inputs $x$ and $y$ and outputs $a$ and $b$ on  Alice's 
and Bob's sides respectively. But this time, they are constrained to have the value $b=y$. Here,
the inputs and outputs are correlated similarly as in the normal PR box situation. However, due to the 
additional constraint $b=y$, the outputs and the inputs of the PR box are now correlated as 
$a\oplus y=x.y$. This implies that we have  $a=(x\oplus 1)y$.
In this case the outputs for a given pair of inputs are given below (see table I).

\begin{table}[ht]
\caption{CTC-assisted PR-box}
\centering
\begin{tabular}{|c|c||c|c||c|}
\hline 
 (x) & (y) &  (a) &  (b) & p(a,b$|$x,y) \\ 
\hline 
0 & 0 & 0 & 0 & 1 \\ 
\hline 
0 & 1 & 1 & 1 & 1 \\
\hline 
1 & 0 & 0 & 0 & 1 \\
1 & 1 & 0 & 1 & 1 \\ 
\hline 
\end{tabular}
\label{table:PR-CTC}
\end{table}

%From the Table 1, it is clear that there is signaling when Alice's input is $0$;,
%Alice can predict with certainty about Bob's input (also output) from her
%output---signaling from Bob to Alice.

From the above tabular representation (Table I) of the inputs and outputs it is clearly evident that there
is a signaling. For an output $a=0$, given an input $x=0$ on Alice's side, she can
tell instantaneously about the input on Bob's side which happens to be $0$. Similarly, for
$a=1$ and $x=0$, Alice can tell that the input on Bob's side is $1$. However, for $a=0$ for an input $x=1$, 
Alice can never infer about Bob's input which can be either $1$ or $0$.
Thus, we see that there is a violation of no signaling principle at least probabilistically.
One can interpret this by saying that because of the kinematic condition, the 
inputs and the outputs on Bob's side are no longer random. This actually leads to signaling.\\

\section{III. Tripartite Case}

In tripartite case, we consider genuine three-party non-local boxes \cite{piro,bar1}. These boxes
are the only full-correlation boxes, for which all one-party and two-party
correlation terms vanish. The input and output correlations for these three types of
boxes are given by
\begin{eqnarray}
a\oplus b\oplus c= x.y\oplus y.z\oplus z.x,
\label{Box-1}
\end{eqnarray}
\begin{eqnarray}
a\oplus b\oplus c= x.y\oplus x.z,
\label{Box-2}
\end{eqnarray}
\begin{eqnarray}
a\oplus b\oplus c= x.y.z,
\label{Box-3}
\end{eqnarray}
where $x$, $y$ and $z$ are the inputs to the box and $a$, $b$ and $c$ are the  corresponding
outputs. The box  (\ref{Box-1}) which violates both Svetlichny inequality
and Mermin inequality \cite{sve,mer}, is known as Svetlichny box and other two boxes (\ref{Box-2},
\ref{Box-3}) are known as Mermin-type boxes, as they violate only Mermin inequality. 
For the sake of convenience we are referring the box \ref{Box-2} and \ref{Box-3} as the Mermin box of Type I and Type II respectively. In this section, we discuss all these genuine three-party non-local boxes and also discuss how these correlations violate no-signaling principle when one or two of the inputs of the boxes have closed time like world line. In other words, these inputs remain same when they go out of the box. 

\subsection{A. Signaling with Svetlichny Box}
First of all in this subsection we will consider the Svetlichny Box whose inputs and outputs are correlated as $a\oplus b\oplus c= x.y\oplus y.z\oplus z.x$.
The probability distribution for Svetlichny box is given by
\begin{eqnarray}
p(a,b,c|x,y,z) &=& \frac{1}{4} ~~~~~~\text{iff $a\oplus b\oplus c= x.y\oplus
y.z\oplus z.x$}, \nonumber \\
&=&0~~~~~~\text{otherwise}.
\end{eqnarray}

Here, we consider the case when the input to the Alice's side has a CTC world line. As in the previous section,   a CTC input gives the same output, i.e, 
\begin{eqnarray}
x=a.
\end{eqnarray}
The Bob's and Charlie's outputs are correlated as
\begin{eqnarray}
b\oplus c = x\cdot y\oplus y\cdot z\oplus z\cdot x\oplus x.
\end{eqnarray}

In this case, when Bob and Charlie share their inputs and outputs  then
there is a signaling from Alice to Bob-Charlie. In other words, in this situation, there is signaling from CTC world  to causality respecting world. When both Bob and Charlie give same
input, i.e., 0 (1) then Alice's input is $b\oplus c$ ($b\oplus c\oplus 1$).
The signaling is probabilistic in the sense that they recover the Alice's input only for four cases and unable to do it for 
remaining four cases,
\begin{eqnarray}
b\oplus c= x\cdot y\oplus y\cdot z \oplus z\cdot x \oplus x, \nonumber \\
b \oplus c=z\oplus x ~~~~~~~\text ({\rm if},~~ y=z).
\label{Cond.1}
\end{eqnarray}
Now, if both Bob and Charlie give the input `0' (i.e., $y=z=0$) then they know
about Alice's input (via, the condition) 
\begin{eqnarray}
x=b\oplus c.
\end{eqnarray}
If they give input `1' (i.e., $y=z=1$) then, they can know about Alice's input by using 
\begin{eqnarray}
x=b\oplus c\oplus 1.
\end{eqnarray} 

If both Bob and Charlie decide that they use different inputs, i.e., $y\neq z$
($y\oplus z=1$ and $y.z=0$) then they can know Alice's input in a simpler way which is
given by
\begin{eqnarray}
x=b\oplus c.
\end{eqnarray}
In the similar manner, one can find out that the signaling takes place when other party's inputs are from the CTC world.

Now, we consider the case when Bob's and  Charlie's inputs are CTC-assisted. This puts restriction on inputs and outputs of Bob and Charlie as follows
\begin{eqnarray}
y=b ,\nonumber \\
z=c.
\end{eqnarray}
Thus Alice's output is correlated as
\begin{eqnarray}
a=x\cdot y\oplus y\cdot z\oplus z\cdot x\oplus y\oplus z.
\end{eqnarray}
The correlation table in presence of CTC inputs at both Bob's and Charlie's
sides is given by Table II.\\
\begin{table}[ht]
\caption{CTC inputs on Bob's and Charlie's sides}
\centering
\begin{tabular}{|c|c|c||c|c|c||c|}
\hline 
 (x) &  (y) &  (z) &  (a) & (b) &  (c) & p(a,b$|$x,y) \\ 
\hline 
0 & 0 & 0 & 0 & 0 & 0 & 1  \\ 
\hline 
0 & 0 & 1 & 1 & 0 & 1 & 1  \\ 
\hline 
0 & 1 & 0 & 1 & 1 & 0 & 1  \\ 
\hline 
1 & 0 & 0 & 0 & 0 & 0 & 1  \\ 
\hline 
0 & 1 & 1 & 1 & 1 & 1 & 1  \\ 
\hline 
1 & 0 & 1 & 0 & 0 & 1 & 1  \\  
\hline 
1 & 1 & 0 & 0 & 1 & 0 & 1  \\  
\hline 
1 & 1 & 1 & 1 & 1 & 1 & 1  \\   
\hline 
\end{tabular} 
\label{table:SPR-BCCTC}
\end{table}

In this case, if Alice's input and output are same, she can correctly infer about
the inputs of both Bob and Charlie without any communication from them. This is  signaling
from both Bob and Charlie to Alice. If Alice's input and output are different, then there
is still signaling. For example, with Alice's input `$x=0$' and her output `$a=1$', she
knows that `$b=0$' and `$c=0$' never happens. Furthermore, if Bob helps her, she can find out
Charlie's input if their inputs are same. However, such a situation is quite impractical because there is 
no feasible way of communicating between CTC and CR (causality respecting) world. Similar arguments holds for other two
cases, i.e., Alice's and Bob's inputs are CTC inputs, or Alice's and Charlie's inputs are
CTC inputs.

\subsection{B. Signaling with Mermin Box Type I}
In this subsection,  we examine three-party non-local Mermin Type I box.  Its outputs and inputs are correlated as $a\oplus b\oplus c= x.y\oplus x.z$.
The corresponding probability distribution in this case is given by
\begin{eqnarray}
p(a,b,c|x,y,z) &=& \frac{1}{4} ~~~~~~\text{iff $a\oplus b\oplus c= x.y\oplus x.z$},
\nonumber \\
&=&0~~~~~~\text{otherwise}.
\end{eqnarray}

First, we consider the case when Alice's input is a CTC input, i.e.,
\begin{eqnarray}
x=a.
\end{eqnarray}
Then Bob's and Charlie's outputs are correlated as
\begin{eqnarray}
b\oplus c &=& x\cdot y\oplus x\cdot z\oplus x, \nonumber \\
&=& x\cdot(y\oplus z\oplus 1).
\label{AMermin2}
\end{eqnarray}

From Eq.(\ref{AMermin2}), it is clearly seen that when Bob and Charlie give same
input (i.e., $y\oplus z=0$) then
\begin{eqnarray}
x=b\oplus c.
\end{eqnarray}
Thus signaling can occur, when Bob and Charlie share information about their inputs and outputs.
When their inputs are anti-correlated, there is no signaling. So,
signaling occurs in two cases among eight cases. This happens when Bob and
Charlie communicate with each other.\\

Now consider the case when Bob's input is a CTC input, i.e.,
\begin{eqnarray}
y=b.
\end{eqnarray}
We also assume that Alice and Charlie can communicate with each other. The
Alice's and Charlie's outputs are correlated as
\begin{eqnarray}
a\oplus c &=& x\cdot y\oplus x\cdot z\oplus y ,\nonumber \\
&=& y\cdot (x\oplus 1) \oplus x.z,  \nonumber \\
y\cdot (x\oplus 1) &=& a\oplus c \oplus x.z.
\label{BMermin2}
\end{eqnarray}
%The correlation table in presence of CTC-assisted box at Alice's side is given by

%\begin{table}[ht]
%\caption{CTC-assisted Alice's box}
%\centering
%\begin{tabular}{|c|c|c|c|c|c|c|}
%\hline 
%x &amp; y &amp; z &amp; a &amp; b &amp; c &amp; Probability \\ 
%\hline 
%0 &amp; 0 &amp; 0 &amp; 0 &amp; 0 &amp; 0 &amp;1/2  \\ 
%  &amp;   &amp;   &amp; 0 &amp; 1 &amp; 1 &amp; 1/2 \\ 
%\hline 
%0 &amp; 0 &amp; 1 &amp; 0 &amp; 0 &amp; 0 &amp;1/2  \\ 
%  &amp;   &amp;   &amp; 0 &amp; 1 &amp; 1 &amp; 1/2 \\ 
%\hline 
%0 &amp; 1 &amp; 0 &amp; 0 &amp; 0 &amp; 0 &amp;1/2  \\ 
%  &amp;   &amp;   &amp; 0 &amp; 1 &amp; 1 &amp; 1/2 \\ 
%\hline 
%1 &amp; 0 &amp; 0 &amp; 1 &amp; 0 &amp; 1 &amp;1/2  \\ 
%  &amp;   &amp;   &amp; 1 &amp; 1 &amp; 0 &amp; 1/2 \\ 
%\hline 
%0 &amp; 1 &amp; 1 &amp; 0 &amp; 0 &amp; 0 &amp;1/2  \\ 
%  &amp;   &amp;   &amp; 0 &amp; 1 &amp; 1 &amp; 1/2 \\ 
%\hline 
%1 &amp; 0 &amp; 1 &amp; 1 &amp; 0 &amp; 0 &amp;1/2  \\  
%  &amp;   &amp;   &amp; 1 &amp; 1 &amp; 1 &amp; 1/2 \\  
%\hline 
%1 &amp; 1 &amp; 0 &amp; 1 &amp; 0 &amp; 0 &amp;1/2  \\  
%  &amp;   &amp;   &amp; 1 &amp; 1 &amp; 1 &amp; 1/2 \\ 
%\hline 
%1 &amp; 1 &amp; 1 &amp; 1 &amp; 0 &amp; 1 &amp;1/2  \\  
%  &amp;   &amp;   &amp; 1 &amp; 1 &amp; 0 &amp; 1/2 \\   
%\hline 
%\end{tabular} 
%\label{table:SPR-ACTC}
%\end{table}\\
From Eq.(\ref{BMermin2}), it can be seen that when Alice's input is `0' then
\begin{eqnarray}
y=a\oplus c.
\end{eqnarray}
Therefore, signaling occurs only when Alice and Charlie discuss about there input and output.
However, when Alice's input is `1' there is no signaling. Here also,
signaling occurs in two cases among eight cases. The other case, when Charlie's input is CTC input then it is
similar to this case.

Next, we consider the case when  Bob's and Charlie's inputs are taken from CTC world , i.e.,
\begin{eqnarray}
y&=&b, \nonumber \\
z&=&c.
\end{eqnarray}
In this case Alice's and Charlie's outputs are correlated as
\begin{eqnarray}
a\oplus c &=& x\cdot y\oplus x\cdot z\oplus y, \nonumber \\
&=& y\cdot (x\oplus 1) \oplus x.z.  \nonumber \\
\label{BCMermin2}
\end{eqnarray}

The correlations in the presence of CTC-assisted box at both Bob's and Charlie's
side are given in Table III.\\
\begin{table}[ht]
\caption{CTC inputs on Bob's and Charlie's side}
\centering
\begin{tabular}{|c|c|c||c|c|c||c|}
\hline 
 (x) &  (y) &  (z) &  (a) &  (b) &  (c) &  p(a,b$|$x,y) \\ 
\hline 
0 & 0 & 0 & 0 & 0 & 0 & 1  \\ 
\hline 
0 & 0 & 1 & 1 & 0 & 1 & 1  \\ 
\hline 
0 & 1 & 0 & 1 & 1 & 0 & 1  \\ 
\hline 
1 & 0 & 0 & 0 & 0 & 0 & 1  \\ 
\hline 
0 & 1 & 1 & 0 & 1 & 1 &1  \\ 
\hline 
1 & 0 & 1 & 0 & 0 & 1 &1  \\  
\hline 
1 & 1 & 0 & 0 & 1 & 0 & 1  \\  
\hline 
1 & 1 & 1 & 0 & 1 & 1 & 1  \\   
\hline 
\end{tabular} 
\label{table:SPR-BCCTC}
\end{table}\\
In this case, Alice cannot find out the inputs of the other two parties without
communication with them, whereas in the Svetlichny case, it is possible in probabilistic
way. When, say, Bob helps Alice then it is possible for her to  know Charlie's
input for all the eight cases. However, such communication between CR and CTC world 
is not considered practical.

There  can be another situation where Alice's and Bob's inputs are CTC inputs, i.e.,
\begin{eqnarray}
x&=&a, \nonumber \\
y&=&b.
\end{eqnarray}

In our case Alice's and Charlie's outputs are correlated as
\begin{eqnarray}
c &=& x\cdot (y\oplus z\oplus 1) \oplus y.
\label{ABMermin2}
\end{eqnarray}

The correlations in the presence of CTC inputs at both Bob's and Alice's
side are given in Table IV.\\
\begin{table}[ht]
\caption{CTC inputs on Alice's and Bob's side}
\centering
\begin{tabular}{|c|c|c||c|c|c||c|}
\hline 
 (x) &  (y) &  (z) &  (a) &  (b) &  (c) &  p(a,b$|$x,y) \\ 
\hline 
0 & 0 & 0 & 0 & 0 & 0 & 1  \\ 
\hline 
0 & 0 & 1 & 0 & 0 & 0 & 1  \\ 
\hline 
0 & 1 & 0 & 0 & 1 & 1 & 1  \\ 
\hline 
1 & 0 & 0 & 1 & 0 & 1 & 1  \\ 
\hline 
0 & 1 & 1 & 0 & 1 & 1 & 1  \\ 
\hline 
1 & 0 & 1 & 1 & 0 & 0 & 1  \\  
\hline 
1 & 1 & 0 & 1 & 1 & 1 & 1  \\  
\hline 
1 & 1 & 1 & 1 & 1 & 0 & 1  \\   
\hline 
\end{tabular} 
\label{table:M2PR-ABCTC}
\end{table}\\
In this scenario, Alice alone cannot find the inputs of one of the two parties without
communication from the other, whereas in Svetlitchny case it is possible in probabilistic
way. When Charlie helps Alice then it is possible for her to  know Bob's input only when Alice's
input is zero (follows from Eq.\ref{ABMermin2}) for all the eight
cases. However, such type of communication between Alice and Charlie is not feasible. Similar conclusions one can draw for the remaining cases.

\subsection{C. Signaling with Mermin Box Type II}
Here we discuss the Type II Mermin box whose correlations for input and output are defined as $a\oplus b\oplus c= x.y.z$.
The probability distribution for this type of correlation is given by
\begin{eqnarray}
p(a,b,c|x,y,z) &=& \frac{1}{4} ~~~~~~\text{iff $a\oplus b\oplus c= x.y.z$}, \nonumber \\
&=&0~~~~~~\text{otherwise}.
\end{eqnarray}

Now once again we consider the cases when one of the party has access to a CTC system.
First of all, we consider the case when Alice has access to the CTC inputs. So, we have
\begin{eqnarray}
x=a.
\end{eqnarray}
Then,  Bob's and Charlie's outputs are correlated as
\begin{eqnarray}
b\oplus c &=& x\cdot (y\cdot z \oplus 1).
\label{AMermin3}
\end{eqnarray}
From Eq.(\ref{AMermin3}), it is clear that if Bob and Charlie discuss
about their inputs and outputs then there is signaling from Alice to Bob-Charlie,
i.e., they probabilistically know Alice's input. The condition for signaling is
\begin{eqnarray}
y\cdot z=0.
\end{eqnarray}
Then we have,
\begin{eqnarray}
x=b\oplus c.
\end{eqnarray}
In the other case, when
\begin{eqnarray}
y\cdot z=1,
\end{eqnarray}
they are unable to know the input of Alice. The similar argument is valid when Bob
and Charlie have access to a CTC system respectively.

Here, we consider the  case when Bob's and Charlie's inputs are CTC inputs, i.e.,
\begin{eqnarray}
y&=&b,\nonumber \\
z&=&c.
\end{eqnarray}
The Alice's output is then given by 
\begin{eqnarray}
a &=& x\cdot y\cdot z \oplus y\oplus z. 
\label{AMermin4}
\end{eqnarray}
From Eq.(\ref{AMermin4}),  it is clear that if Alice and Bob share 
their inputs and outputs, then there is signaling from Charlie to Alice-Bob,
i.e., they probabilistically know Charlie's input. The condition for signaling is
\begin{eqnarray}
y=0 ~~~~~~ (x=0).
\end{eqnarray}
Then we have,
\begin{eqnarray}
z=a~~~~~~(z=a\oplus y).
\end{eqnarray}
However, we are restricted to a situation where no such discussion is possible between
CTC and CR worlds. 
The similar argument is valid when both Alice's  and Charlie's inputs and Alice' and
Bob's inputs are CTC inputs respectively.\\

\section{IV. Conclusion}
The basic objective of this work was to the study the power of non-local boxes in the presence of CTC. It is interesting to see that the existing no-signaling boxes shows s
ignaling in the presence of  CTCs. 
Here, we have shown that the PR box correlations  and their generalizations to tripartite cases can  violate
signaling in the presence of the closed time-like curves. Some time it is direct violation as the signaling takes 
place without any communication and sometimes it is indirect violation as the signaling takes place as a result of 
communication. Our results once
again show the capability of CTC in making a no-signalling theory a
signalling one. It would be interesting to explore
if another model of the CTC qubits by Bennett and others also transforms a no-signally box to a 
signaling box.

\noindent
\textit {Acknowledgment:} The authors acknowledge useful discussions with Dagomir Kaszlikowski and
Tomasz Paterek.  T. Pramanik thanks UGC, India for financial support.


\begin{thebibliography}{100}
\bibitem{woo} W. K. Wootters, Phys. Rev. Lett. {\bf 80}, 2245 (1998).  


\bibitem{pop} S. Popescu, D. Rohrlich, Found. Phys. {\bf 24}, 379 (1994).

\bibitem{cri} B. S Cirel'son, Lett. Math. Phys. {\bf 4}, 93 (1980).

\bibitem{bar} J. Barrett, S. Pironio, Phys. Rev. Lett. {\bf 95}, 140401 (2005).

\bibitem{marco} S. Marcovitch, B. Reznik, L. Vaidman, Phys. Rev. A {\bf 75}, 022102 (2007).

%\bibitem{mat} M. Barnett, F. Dowker, D. Rideout, J. Phys. A: Math. Theor. {\bf 40},  7255 (2007).

\bibitem{fit} M. Fitzi, E. Hanggi, V. Scarani, S. Wolf, J. Phys. A: Math. Theor. {\bf 43}, 465305 (2010).

\bibitem{piro} S. Pironio, J.-Daniel Bancal and V. Scarani, J. Phys. A: Math. Theor. \textbf{44}  065303 (2011).

\bibitem{bar1} J. Barrett, N. Linden, S. Massar, S. Pironio, S. Popescu, and D. Roberts, Phy. Rev. A \textbf{71},
022101 (2005))

\bibitem{sve} G. Svetlichny, Phys. Rev. D \textbf{35}, 3066 (1987) 

\bibitem{mer} N. D. Mermin, Phys. Rev. Lett. \textbf{65}, 15 (1990).

\bibitem{tas} D. S. Tasca, S. P. Walborn, F. Toscano, P. H. Souto Ribeiro, Phys. Rev A. {\bf 80}, 030101(R) (2009). 

\bibitem{scar} V. Scarani, AIP. Conf. Proc.  {\bf 844},  309 (Melville, New York, 2006).

%\bibitem{kunkri} S. Kunkri and S. K. Choudhary, Phys. Rev. A {\bf 72}, 022348 (2005).

\bibitem{ahanj} A. Ahanj, P. S. Joag, Simulation of partial entanglement with one cbit and one M-box,
arXiv:0807.214 (2008).

\bibitem{ahanj1} A. Ahanj, P. S. Joag, Simulation of a partially entangled two qubit state correlation with one PR-Box and one M-box, 
arXiv:1104.2491 (2011).

\bibitem{kunkri1} S. Kunkri, G. Kar, S. Ghosh, A. Roy, Quant. Inf. Comp.  {\bf 7}, 319 (2007). 

\bibitem{deu} D. Deutsch, Phys. Rev. D. {\bf 44}, 3197 (1991).

\bibitem{bru} T. A. Brun, Found. Phys. Lett. {\bf 16}, 245 (2003).
% arXiv:gr-qc/0209061v1.

\bibitem{bac} D. Bacon, Phys. Rev. A. {\bf 70}, 032309 (2004).
%arXiv:quant-ph/0309189v3.

\bibitem{aar} S. Aaronson, J. Watrous, Proc. R. Soc. A. {\bf 465}, 631
(2009).

\bibitem{ahn} D. Ahn, T. C. Ralph, R. B. Mann, Any quantum state can be cloned in the presence of closed timelike curves, 
arXiv:1008.0221  (2010). 
% arXiv:0808.2669v1.

\bibitem{bru1} T. A. Brun, J. Harrington, M. M. Wilde, Phys. Rev.
Lett. {\bf 102}, 210402 (2009). 
%arXiv:0811.1209v2.

\bibitem{pati} A. K. Pati, I. Chakrabarty, P. Agrawal, Phys. Rev. A \textbf{84}, 062325 (2011).

\bibitem{ben1} C. H. Bennett, D. Leung, G. Smith, J. A. Smolin, 
Phys. Rev. Lett. {\bf 103}, 170502 (2009).





% arXiv:0908.3023v1.

\end{thebibliography}
\end{document}